\documentclass{aa} 
\usepackage{graphicx}
\usepackage{txfonts}
\usepackage{natbib}
\usepackage{verbatim}
\usepackage{longtable}
\usepackage{lscape}
\usepackage{afterpage}
\usepackage{multirow}
\bibpunct{(}{)}{;}{a}{}{,}

\begin{document}

\title{Theoretical evaluation of PAH dication properties}

\author{G. Malloci\inst{1,2}
	\and
	C. Joblin\inst{1}
	\and
	G. Mulas\inst{2}
}

\institute{
Centre d'Etude Spatiale des Rayonnements, CNRS et Universit\'e Paul 
Sabatier Toulouse~3, Observatoire Midi-Pyr\'en\'ees, 9 Avenue du Colonel Roche, 
31028 Toulouse cedex 04 (France)
\textemdash\email{[giuliano.malloci; christine.joblin]@cesr.fr}
\and
Istituto Nazionale di Astrofisica~\textendash~Osservatorio Astronomico di Cagliari, 
Strada n.54, Loc. Poggio dei Pini, I\textendash09012 Capoterra (CA) (Italy)
\textemdash\email{[gmalloci; gmulas]@ca.astro.it}
}
\date{Received 18 July 2006; 25 September 2006}

\abstract
{\emph{Aims}.
We present a systematic theoretical study on 40~polycyclic aromatic 
hydrocarbons dications (PAHs$^{++}$) containing up to 66~carbon atoms.\\
{\emph{Methods}.
We performed our calculations using well established quantum\textendash chemical 
techniques in the framework of the density functional theory (DFT) to obtain 
the electronic ground\textendash state properties, and of the time\textendash dependent DFT 
(TD\textendash DFT) to evaluate the excited\textendash state properties.}\\
{\emph{Results}.
For each PAH$^{++}$ considered we computed the absolute visible\textendash UV 
photo\textendash absorption cross\textendash section up to about 30~eV.
We also evaluated their vibrational properties and compared them to those of 
the corresponding neutral and singly\textendash ionised species. We estimated the 
adiabatic and vertical second ionisation energy $\Delta I$ through total energy 
differences.}\\
{\emph{Conclusions}.
The $\Delta I$ values obtained fall in the energy range 8\textendash13~eV, confirming that 
PAHs could reach the doubly\textendash ionised state in HI regions. 
The total integrated IR absorption cross\textendash sections show a marked increase upon 
ionization, being on the average about two and five times larger for PAHs$^{++}$ 
than for PAHs$^+$ and PAHs, respectively.
The visible\textendash UV photo\textendash absorption cross\textendash sections for the 0, +1 and +2 
charge\textendash states show comparable features but PAHs$^{++}$ are found to absorb 
slightly less than their parent neutrals and singly ionized species between 
$\sim7$ and $\sim12$~eV.
Combining these pieces of information we found that PAHs$^{++}$ should
actually be more stable against photodissociation  than PAHs and PAHs$^+$, 
\emph{if} dissociation tresholds are nearly unchanged by ionization.}\\

\keywords{astrochemistry \textemdash{} molecular data \textemdash{} molecular processes 
\textemdash{} ISM: molecules \textemdash{} ultraviolet: ISM \textemdash{} 
infrared: ISM}}

\authorrunning{Malloci et al.}
\titlerunning{Theoretical evaluation of PAH dication properties}

\maketitle

\section{Introduction} \label{introduction}

Several spectroscopic features, observed in absorption or in emission in a 
large number of interstellar environments, are hypothesised to provide 
strong evidence for the presence of large carbon\textendash based molecules 
\citep{ehr00}. Among them are the diffuse 
interstellar bands \citep[DIBs,][]{her95,ful00}, seen in absorption in the 
visible\textendash UV; the aromatic infrared bands \citep[AIBs,][]{pee04,tie05}, 
dominating the near and mid\textendash IR spectrum of many interstellar sources; the 
extended red emission \citep[ERE,][]{wit04,wit06} and the 
blue\textendash luminescence \citep[BL,][]{vij04,vij05b,vij05a}, wide emission bands 
peaking in the red and blue parts of the visible spectrum, respectively; 
the UV\textendash bump at $\sim220$~nm and the far\textendash UV rise in the general interstellar 
extinction curve \citep[]{dra03,fit04}. 

Based on their spectral properties, their high photostability and their
organic nature, polycyclic aromatic hydrocarbons (PAHs) are believed to be 
abundant in the interstellar medium \citep[ISM,][]{pug89,all89}. Isolated
gas\textendash phase PAHs absorb efficiently visible\textendash UV radiation and are therefore 
likely to contribute to the DIBs spectrum \citep{leg85,van85,cra85}
and to the UV\textendash bump and far\textendash UV rise of the extinction curve 
\citep{job92a,job92b}. After photon absorption, they undergo a fast 
radiation-less transition (internal conversion or inter-system crossing), 
and subsequently relax radiatively by electronic fluorescence and/or 
phosphorescence (possibly contributing to BL and ERE) and IR emission in 
their active C\textendash C and C\textendash H vibrational modes 
\citep[hence producing AIBs,][]{leg84,all85}.

Interstellar PAHs are expected to exist in different charge and hydrogenation 
states depending on the physical conditions of the host environment 
\citep{tie05}. In particular, the possible presence of PAH dications 
(PAHs$^{++}$) in the ISM, first proposed by \citet{lea86}, has recently received 
further interest, following the proposal that PAHs$^{++}$ could be the carriers
of ERE \citep{wit06}. Theoretical works on the IR properties 
of a few PAHs$^{++}$ \citep{ell99,bau00,pau02} show that their vibrational 
patterns are similar to those of their parent singly ionized species (PAHs$^+$).
From modelling studies it is predicted that specific values of electron 
density and radiation field intensity ought to lead to a population of 
PAHs$^{++}$ in excess of that of PAHs$^+$ \citep{bak01b,bak01a}. Therefore 
PAHs$^{++}$ would also contribute to the AIBs.

While the first ionisation energies $I^+$ for more than 200~PAHs containing up 
to 48 carbon atoms are well documented \citep{eil81}, only a few 
experimental determinations of the double ionisation energy $I^{++}$ of some
PAHs have been published to date. Using photon impact, electron impact and 
charge stripping techniques, \citet{tob94} performed a comprehensive study of 
$I^+$ and $I^{++}$ for 21 different PAHs up to the size of coronene 
(C$_{24}$H$_{12}$). More recently, \citet{sch01} presented a combination between 
photo\textendash ionisation, charge stripping, reactivity studies and quantum\textendash chemical 
calculations to obtain the single and double ionisation energies of 
corannulene (C$_{20}$H$_{10}$) and coronene. 

From the known values of $I^+$ and $I^{++}$, it was suggested 
\citep{lea86,lea87,lea89,lea95b,lea96} that PAHs$^{++}$ could be formed in 
diffuse HI regions.
The proposed formation mechanism is a sequential two\textendash photons absorption 
process in which the ionisation of a PAH is followed by a further ionisation 
of PAH$^+$. The minimum energy needed for the first step is the single 
ionisation energy $I^+$, while the second step requires a photon energy of at 
least $\Delta I$ = $I^{++} - I^{+}$; the maximum amount of internal energy in the 
newly formed PAHs$^{++}$ is therefore 13.6 eV - $\Delta I$. 
Since the total energy of
the dication is larger than the sum of the total energies of two singly 
ionised fragments, PAHs$^{++}$ are metastable \citep{ros04} and could be 
dissociated via Coulomb explosion as shown by experiments 
\citep{lea89a,lea89b}. 
However, theoretical calculations on the fragmentation pathways of the benzene 
dication \citep{ros04} show that this charge separation 
mechanism should not be a significant channel in most astrophysical 
environments. This is consistent with the observed production of significant 
amounts of PAHs$^{++}$ in soft ionisation experiments \citep{led99}.
Therefore PAHs$^{++}$ could be found in the ISM. 

To evaluate the possible contribution of PAHs$^{++}$ to ERE 
\citet{wit06} 
computed the first low\textendash lying electronic transitions of five representative 
PAHs$^{++}$ up to the size of ovalene (C$_{32}$H$_{14}$), showing that they fall 
indeed in the spectral range of the observed ERE. 
Laboratory measurements of the yield of optical 
fluorescence by PAHs$^{++}$ are however needed to test this hypothesis. To the 
best of our knowledge, with the exception of the work by \citet{wit06}, no 
other electronic spectra of PAHs$^{++}$ have been published to date. In 
particular, a detailed study of the electronic excitation properties 
for a large sample of PAHs$^{++}$ up to the energies excitable in typical 
interstellar sources, together with their vibrational analyses, has been 
missing until now.

In the last few years we have started a long\textendash term project to produce an atlas
of synthetic absorption and emission spectra of specific PAHs, to be compared 
with astronomical observations \citep{job02,mal03,mul03,mul06b,mul06a,mul06c}. 
This requires the use of some key molecular parameters of PAHs as a basis to 
run Monte\textendash Carlo simulations \citep{bar83,mul98,job02} of their photophysics 
in the ISM. 
We report here these fundamental data for a sample of 40~PAHs$^{++}$ ranging in 
size from naphthalene and azulene (C$_{10}$H$_{8}$) to circumovalene 
(C$_{66}$H$_{20}$). Even if PAHs and related species containing from less than 50 
to more than 200 carbon atoms are expected to be present in the ISM 
\citep{bou99}, we here restricted ourselves to species containing up to a 
maximum of 66~carbon atoms, since computational costs scale steeply with 
dimensions. 
Following our previous work on neutral, cationic and anionic PAHs 
\citep{mal04,mal05}, we computed the absolute photo\textendash absorption cross\textendash section 
for the doubly\textendash ionised species. For these same molecules we also evaluated 
the IR spectral properties extending the sample of specific PAH$^{++}$ for which 
they were previously available \citep{ell99,bau00,pau02}. New calculations, 
together with our previous results at the same level of theory for the 
corresponding PAHs and PAHs$^+$, enabled us to estimate, via total energy 
differences, the adiabatic and vertical $I^+$ and $I^{++}$.
Although the computational approach we used for this purpose is not sufficient 
to predict absolute ionisation energies within chemical accuracy ($\pm$0.1~eV), 
it is in reasonable agreement with the available experimental data and can be 
easily extended to larger PAHs in the near future.

Section~\ref{computational} contains the technical details for the 
calculation of the ionisation energies and the vibrational properties
(Sect.~\ref{dft}) and the photo\textendash absorption spectra (Sect.~\ref{tddft}). 
The results obtained are presented in Sect.~\ref{results} and discussed 
in Sect.~\ref{discussion}. Our conclusions are reported in 
Sect.~\ref{conclusions}.

\section{Computational details} \label{computational}

Due to the large number of electrons in the molecules considered 
(416 in the largest one, circumovalene), an \emph{ab initio} study based on 
the direct solution of the many\textendash electron Schr\"odinger equation currently has 
prohibitive computational costs. We here used the density functional theory 
\citep[DFT,][]{jon89} and its time\textendash dependent extension 
\citep[TD\textendash DFT,][]{mar04}, which are the methods of choice for the 
study, respectively, of the ground\textendash state and the excited\textendash state properties of 
such complex molecules as PAHs.
 
\subsection{Ionisation energies and vibrational properties}
\label{dft}

Adiabatic single and double ionisation energies (AIEs) are evaluated as the 
difference between the total energies of the neutral and the corresponding 
cation and dication in their most stable configurations, respectively. These 
quantities, therefore, take into account the structural relaxation of the 
molecule after each ionisation process. The single and double vertical 
ionisation energies (VIEs), computed at the optimised geometry of the neutral 
molecule, neglect structural relaxation but include wave-functions relaxation 
following the removal of one and two electrons. To estimate AIEs and VIEs 
we used the hybrid B3LYP functional, a combination of Becke's three 
parameters exchange functional \citep{bec93} and of the Lee\textendash Yang\textendash Parr 
gradient\textendash corrected correlation functional \citep{lee88}. Hybrid DFT 
functionals are a little more expensive than other exchange\textendash correlation 
functional but yield better results. The B3LYP functional, in particular, is 
widely used in the study of PAHs and related species \citep[][]{mar96,lan96,
wib97,bau97,kat99,kat02,des00,bau00,bau01,sch01,rie01,bau02,par03,hir03,del03,
ros04,woo04,jol05,wit06,kad06}. To assess the impact of this choice we also 
considered some more recently developed functionals, such as the so\textendash called 
B\textendash97 \citep{bec97} and HCTH \citep{ham98}, but no significants improvements, 
if any, were obtained with respect to B3LYP for the calculation of the 
ionisation energies of PAHs. 

For this part of the work we used the Gaussian\textendash based DFT module of the
\textsc{NWChem} code \citep{apr05}. Geometry optimisations 
were performed using the relatively inexpensive \mbox{4\textendash31G} basis set, 
followed by the full vibrational analyses to confirm the geometries obtained 
to be global minima on the potential energy surface and to evaluate 
zero\textendash point\textendash energy (ZPE) corrections. In the study of the vibrational
properties of PAHs, the combination \mbox{B3LYP/4\textendash31G} was proven to give 
good results scaling all frequencies by an empirical scale\textendash factor 
\citep{lan96,bau97,bau00,bau01,bau02}. We thus adopted the same scaling 
procedure derived by these authors to obtain the IR spectra of all the 
PAHs under study in their 0, +1 and +2 charge states. Since for  open\textendash shell 
systems analytical second derivatives are not yet implemented in the currently 
available version of \textsc{NWChem} (4.7), we performed the vibrational 
analyses for all the cations using the \textsc{Gaussian03} quantum chemistry 
package \citep{g03}.

We then started from the optimal geometries and the 
corresponding self\textendash consistent\textendash field solution obtained in the previous step 
to refine the optimisation with the \mbox{6\textendash31+G$^\star$} basis, a valence double 
zeta set augmented with $d$ polarisation functions and $s$ and $p$ diffuse 
functions for each carbon atom \citep{fri84}. 
Although basis set convergence is not yet expected at this level, in view of 
the large systems under study some compromise between accuracy and 
computational costs had to be made. Thus, while the \mbox{B3LYP/6\textendash31+G$^\star$} 
level of theory gives a good agreement with experiment for the electron 
affinities of PAHs \citep{des00,mal05}, from previous works 
\citep{sch01,del03,woo04,kad06} it is clear that this same level is certainly 
unable to predict absolute ionisation energies within chemical accuracy 
($\pm$0.1eV). For example, using the \mbox{6-311g$^{\star\star}$} basis set, \citet{sch01} 
obtained mean deviations of -0.3 and -0.8~eV by comparing their computed values
of the single and double AIEs of corannulene (C$_{20}$H$_{10}$) and coronene 
(C$_{24}$H$_{12}$) with their measured photoionisation thresholds. Analogously, 
the first AIE of five PAHs in the oligoacenes series from naphthalene 
to hexacene (C$_{4n+2}$H$_{2n+4}$ with n=2, 3, 4, 5, 6), computed by \citet{kad06}
using an even larger
basis set, underestimate the experimental 
values by $0.2\textendash0.4$~eV; the authors concluded that a better description of 
electron correlation is needed to reproduce experimental IEs.

To estimate the effect of basis
set incompleteness, we performed some reference computations for the AIEs of 
naphthalene (C$_{10}$H$_{8}$) and fluorene (C$_{13}$H$_{10}$), using the correlation 
consistent polarised valence basis sets cc\textendash pVDZ and cc\textendash pVTZ  
\citep{dun89}. These bases have been used in a benchmark\textendash quality study of the 
ionisation energies of benzene and oligoacenes, to obtain the impressive 
accuracy of 0.02\textendash0.07~eV \citep{del03}. Even if the cc-pVDZ set is less 
accurate than the 6-31+G* set and the cc-pVTZ is far from complete, this 
comparison provides us some insight into basis set requirements.
As to zero\textendash point\textendash energy corrections, these same 
authors found that typical values for the PAHs considered are in the range 
$0.01\textendash0.03$. Our B3LYP/4-31G values were computed to be of the same order of 
magnitude and were therefore omitted in the evaluation of the IEs, since they 
are smaller than the accuracy of the method used. Although the basis set limit 
is not reached in Table~\ref{basis} \citep[e.g.,][ report 7.88~eV for the 
first AIE of naphthalene]{kad06}, it is clearly seen that the performance of 
the \mbox{6\textendash31+G$^\star$} basis is intermediate between those of Dunning's two 
basis sets. 
We thus expect our theoretical predictions to systematically 
underestimate single and double AIEs roughly by 4 and 5~\% respectively, which 
is sufficiently accurate for the purposes of this work.

\begin{table}
\begin{center}
\caption{\label{basis}
Effect of basis set incompleteness on the evaluation of single and
double AIEs of naphthalene and fluorene at the 
B3LYP level of theory. The reported values (in eV) do not include 
ZPE corrections.}
\begin{tabular}{ccccc}
\hline \hline
\noalign{\smallskip}
Basis & \multicolumn{2}{c}{Naphthalene (C$_{10}$H$_8$)} & 
\multicolumn{2}{c}{Fluorene (C$_{13}$H$_{10}$)}\\
set &  single IE & double IE & single IE & double IE\\
\noalign{\smallskip}
\hline
\noalign{\smallskip}
 4\textendash31g & 7.61 & 20.75 & 7.38 & 19.72 \\
\noalign{\smallskip}
\noalign{\smallskip}
 6\textendash31+g$^\star$ & 7.80$^{a,b}$ & 20.99$^b$ & 7.56 & 19.96 \\
\noalign{\smallskip}
\noalign{\smallskip}
cc\textendash pvdz & 7.75$^c$ & 20.95 & 7.51 & 19.89\\
\noalign{\smallskip}
\noalign{\smallskip}
 cc\textendash pvtz & 7.85$^c$ & 21.09 & 7.61 & 20.04\\
\noalign{\smallskip}
\hline
\noalign{\smallskip}
Exp. & 8.144$\pm$0.001$^{d}$ & 21.5$\pm$0.2$^{e}$ & 7.91$\pm$0.02$^{d}$ & 
21.0$\pm$0.2$^{e}$\\
 \noalign{\smallskip}
\hline
\end{tabular}
\end{center}
$^\mathrm{a}$ The use of the 6-31+G$^{\star\star}$ basis set (including 
polarisation 
functions on both C and H atoms) leads to the same value \citep{woo04};\\
$^\mathrm{b}$ Using the 6-311G$^{\star\star}$ basis set (which is supplemented with a 
third layer of valence functions and includes polarisation functions) 
\citet{sch01} obtain 7.82 and 21.04, respectively;\\
$^\mathrm{c}$ At the same level, \citet{del03} report 7.75 and 7.83~eV;\\
$^\mathrm{d}$ Recommended values \citep[NIST Chemistry WebBook,][]{lia05}.\\
$^\mathrm{e}$ Derived from photon impact experiments \citep{tob94}.
\end{table}

The optimised ground\textendash state structures were all planar, with the exception of 
acenaphthene, fluorene and corannulene. Whenever possible, molecular symmetry 
was assumed during the calculations. In the case of the more symmetric 
molecules considered, symmetry breaking was observed upon single and double
ionisation, as expected from Jahn\textendash Teller distortion \citep{tor99,kat99}.
More specifically, we obtained a symmetry reduction from D$_\mathrm{6h}$ to 
D$_\mathrm{2h}$ for coronene, circumcoronene and hexabenzocoronene, from  
D$_\mathrm{3h}$ to D$_\mathrm{2v}$ for triphenylene and from D$_\mathrm{5v}$ to 
C$_\mathrm{s}$ for corannulene. The complete set 
of the 40 molecules considered and their symmetry point groups are reported in 
Table~\ref{mols}. 

\begin{table}
\begin{center}
\caption{\label{mols}
Complete set of different PAHs considered in this work.}
\begin{tabular}{lc}
\hline \hline
\noalign{\smallskip}
Name (Formula) & Symmetry group \\
\noalign{\smallskip}
\hline
\noalign{\smallskip}
Azulene (C$_{10}$H$_{8}$) & D$_\mathrm{2h}$\\
Naphthalene (C$_{10}$H$_{8}$) &  D$_\mathrm{2h}$\\
Acenaphthylene (C$_{12}$H$_{8}$)& C$_\mathrm{2v}$\\
Biphenylene (C$_{12}$H$_{8}$) & D$_\mathrm{2h}$ \\
Acenaphthene (C$_{12}$H$_{10}$)& C$_\mathrm{2v}$\\
Fluorene (C$_{13}$H$_{10}$)& C$_\mathrm{2v}$\\
Anthracene (C$_{14}$H$_{10}$) & D$_\mathrm{2h}$\\
Phenanthrene (C$_{14}$H$_{10}$)& C$_\mathrm{2v}$\\
Pyrene (C$_{16}$H$_{10}$) & D$_\mathrm{2h}$\\
Tetracene (C$_{18}$H$_{12}$) & D$_\mathrm{2h}$\\
Chrysene (C$_{18}$H$_{12}$)& C$_\mathrm{2h}$\\
Triphenylene (C$_{18}$H$_{12}$)& D$_\mathrm{3h}$\\
Benzo[a]anthracene (C$_{18}$H$_{12}$)& C$_\mathrm{s}$\\
Corannulene (C$_{20}$H$_{10}$)& C$_\mathrm{5v}$\\
Benzo[a]pyrene (C$_{20}$H$_{12}$)& C$_\mathrm{s}$\\
Benzo[e]pyrene (C$_{20}$H$_{12}$)& C$_\mathrm{2v}$\\
Perylene (C$_{20}$H$_{12}$) & D$_\mathrm{2h}$\\
Anthanthrene (C$_{22}$H$_{12}$)& C$_\mathrm{2h}$\\
Benzo[g,h,i]perylene (C$_{22}$H$_{12}$) & D$_\mathrm{2h}$\\
Pentacene (C$_{22}$H$_{14}$) & D$_\mathrm{2h}$\\
Coronene (C$_{24}$H$_{12}$) & D$_\mathrm{6h}$\\
Dibenzo[b,def]chrysene (C$_{24}$H$_{14}$) & C$_\mathrm{2h}$\\
Dibenzo[cd,lm]perylene (C$_{26}$H$_{14}$) & D$_\mathrm{2h}$\\
Hexacene (C$_{26}$H$_{16}$) & D$_\mathrm{2h}$\\ 
Bisanthene (C$_{28}$H$_{14}$) & D$_\mathrm{2h}$\\
Benzo[a]coronene (C$_{28}$H$_{14}$) & D$_\mathrm{2h}$\\
Dibenzo[bc,kl]coronene (C$_{30}$H$_{14}$) & D$_\mathrm{2h}$\\
Dibenzo[bc,ef]coronene (C$_{30}$H$_{14}$) & C$_\mathrm{2v}$\\
Terrylene (C$_{30}$H$_{16}$) & D$_\mathrm{2h}$\\
Ovalene (C$_{32}$H$_{14}$) & D$_\mathrm{2h}$\\
Tetrabezo[bc,ef,kl,no]coronene (C$_{36}$H$_{16}$) & D$_\mathrm{2h}$\\
Circumbiphenyl (C$_{38}$H$_{16}$) & D$_\mathrm{2h}$\\
Circumanthracene (C$_{40}$H$_{16}$) & D$_\mathrm{2h}$\\
Quaterrylene (C$_{40}$H$_{20}$) & D$_\mathrm{2h}$\\
Circumpyrene (C$_{42}$H$_{16}$) & D$_\mathrm{2h}$\\
Hexabenzocoronene (C$_{42}$H$_{18}$)& D$_\mathrm{6h}$\\
Dicoronylene (C$_{48}$H$_{20}$) & D$_\mathrm{2h}$\\
Pentarylene (C$_{50}$H$_{8}$) & D$_\mathrm{2h}$\\
Circumcoronene (C$_{54}$H$_{18}$)& C$_\mathrm{6h}$\\
Circumovalene (C$_{66}$H$_{20}$) & D$_\mathrm{2h}$\\
\hline
\end{tabular}
\end{center}
\end{table}

\subsection{The real\textendash time TD\textendash DFT method applied to PAHs$^{++}$}
\label{tddft}

TD\textendash DFT calculations were shown to be a powerful tool to calculate electronic 
excitation properties for neutral PAHs \citep{hei00,par03,kad06} as well as 
radical ions up to large species 
\citep{hir99,hir03,hal00,hal03,wei01,wei03,wei05,wit06}. 
In our previous studies \citep{mal04,mal05} we calculated the photo\textendash absorption
cross\textendash section $\sigma(E)$ of neutral and singly\textendash ionised PAHs using the real\textendash time 
real\textendash space TD\textendash DFT implementation of the \textsc{octopus} computer 
code \citep{mar03}. We performed these calculations using the widely used 
local-density approximation (LDA), specifically with the exchange-correlation 
energy density of the homogeneous electron gas \citep{cep80} parametrised by 
\citet{per81}. The comparison with the experimental data available for 
a few neutral PAHs \citep{job92a,job92b} shows that our results reproduce 
the overall far\textendash UV behaviour up to about 30~eV. This includes the broad 
absorption peak dominated by $\sigma^*\gets\sigma$ transitions, which matches 
well both in position and width. Concerning the low\textendash lying excited states of 
$\pi^*\gets\pi$ character occurring in the near\textendash IR, visible and near\textendash UV spectral 
range, the computed vertical excitation energies are precise to within a few 
tenths of eV (in the range 0.1-0.4 eV) which are indeed the typical accuracies 
achievable by TD\textendash DFT using hybrid or gradient\textendash corrected exchange\textendash correlation 
functionals \citep{hir99,hir03}. We proved these data 
to be reliable enough for detailed modelling of PAHs in different charge 
states when laboratory data are missing \citep{mul06a,mul06b,mul06c}. Starting 
from the ground\textendash state geometries obtained as described in the previous 
section, we extended here similar calculations to PAHs$^{++}$. 

In the scheme we used the wave-functions are represented by their discretised 
values on a uniform spatial grid. The molecule is perturbed by an impulsive 
electromagnetic field and the time\textendash dependent Kohn\textendash Sham equations are solved 
in real time. From the knowledge of the time\textendash dependent induced dipole moment,
$\sigma(E)$ directly follows by Fourier transform
\footnote{For further details the reader 
is referred to the \textsc{octopus} web site 
\mbox{\texttt{http://www.tddft.org/programs/octopus}}.}. 
The calculations we performed for PAHs$^{++}$ are almost the same as in 
\citet{mal04}, where a thorough description can be found. As already done for 
anions \citep{mal05}, we added an absorbing boundary, which quenches 
spurious resonances due to standing waves in the finite simulation box 
\citep{yab99,mar03}. 

The photo\textendash absorption spectra of a few PAHs$^{++}$ reported by \citet{wit06} were
obtained with a different implementation of TD\textendash DFT, which is based on the 
identification of the poles of the linear response function \citep{cas95}. 
In this latter scheme, computational costs scale steeply with the number of 
required transitions, and the excitation spectra are limited to the 
low\textendash energy part of the spectrum \citep[e.g.,][]{hir99,hir03}. The a
dvantages of the real\textendash time propagation method used here are discussed by 
\citet{lop05}. From the astrophysical point of view, the main step forward we 
achieved using this approach lies in the spectral range covered, that extends 
up to the energies excitable in a typical interstellar source. At the same 
time, the main drawback is that we do not obtain independent information for 
each excited electronic state, such as its symmetry and its description in 
terms of promotion of electrons in a one\textendash electron picture.

\section{Results} \label{results}

All neutral and singly\textendash ionised species were computed as singlet and doublet, 
respectively, while for dications we computed both their singlet and triplet 
ground states. This enabled us to predict the relative energies of electronic 
states having different multiplicities. For almost all of the molecules under 
study (34 out of a total of 40), our calculations predict the dication overall
ground\textendash state to be the singlet. In the other cases, the dications of 
acenaphtylene, triphenylene, corannulene, coronene, hexabenzocoronene and 
circumcoronene, the predicted energy difference between the singlet 
and triplet state is in the range 0.03\textendash0.36~eV, which cannot unambiguously 
identify the ground state, given the accuracy of the method we used.
In this respect DFT\textendash based methods are known to be limited in the evaluation 
of accurate energetic ordering of very close electronic states having 
different multiplicities \citep{sch01}. In addition, the B3LYP functional is 
thought to be biased towards higher spin states \citep{bau00} and better 
levels of wave function favour the singlet state relative to the triplet 
\citep{ros04}. 

\subsection{Single and double ionisation energies}

For each species we derived single and double AIEs, respectively 
as total energy differences between the geometry optimised cations and 
dications and the total energy of the geometry optimised neutral. The 
corresponding VIEs were evaluated as the differences between the single point 
energy of the cation and dication at the optimised neutral geometry and the 
total energy of the optimised neutral. The adiabatic second ionisation energy 
$\Delta I$ is simply given by $I^{++}_\mathrm{ad} \textendash{} I^+_\mathrm{ad}$ while its vertical 
value is obtained through the difference between the single point energy of 
the dication at the optimised cation geometry and the total energy of the 
optimised cation. These results are listed in the Table~\ref{double} of the 
Appendix and sketched in Fig.~\ref{trend} as a function of the number of 
carbon atoms in the molecule. The differences between vertical and adiabatic 
ionisation energies are found to decrease as the size of the molecule 
increases. In all cases however, as already found by \citet{sch01}, these 
values are relatively small, being of the order of 0.1~eV for first and second 
IEs and falling in the range 0.1\textendash0.5~eV for double IEs. 

\begin{figure}
\includegraphics[width=\hsize]{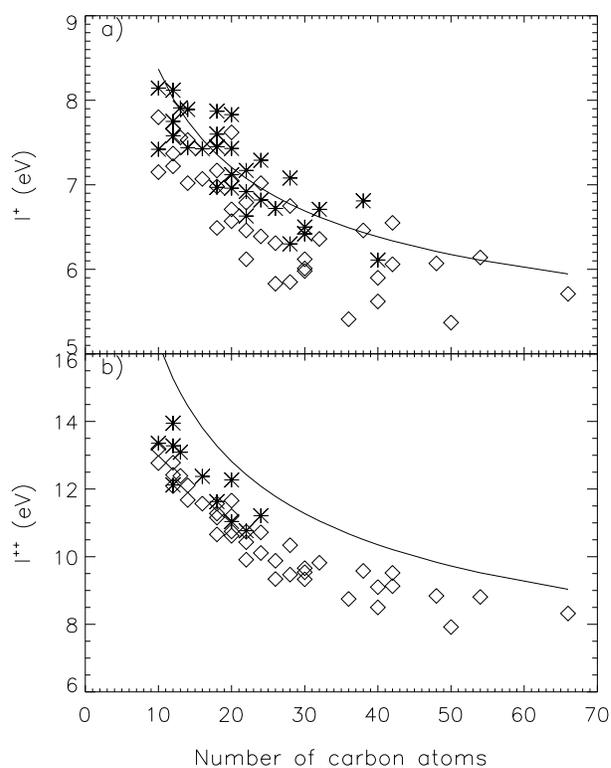}
\caption{Comparison between experimental (asterisks) and computed (diamonds)
first (a) and second (b) AIEs for the 40 PAHs considered in this work as a 
function of the number of carbon atoms in the molecule. The continuous line 
is an empirical estimate \citep[][ p.~191]{tie05}.}
\label{trend}

\end{figure}

\subsection{IR spectral properties}

\begin{figure}
\includegraphics[width=\hsize]{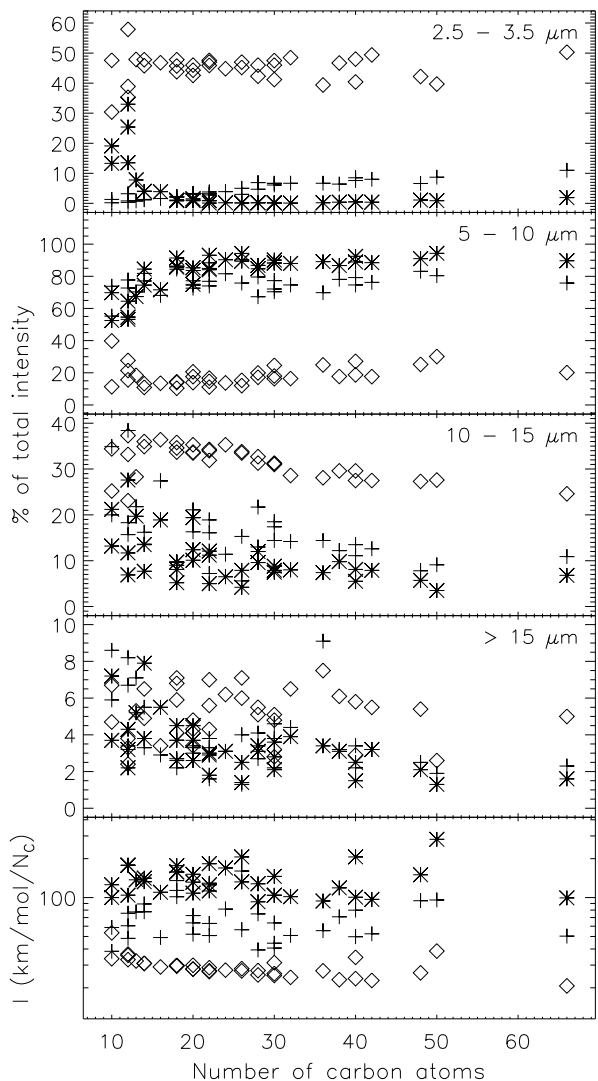}
\caption{Comparison between the fraction of the total integrated IR absorption 
cross\textendash section in the spectral ranges $0\textendash5$, $5\textendash10$, $10\textendash15$~$\mu$m
and $>15$~$\mu$m for the neutrals (diamonds), cations (crosses) and dications 
(asterisks) of the PAHs considered as a function of the number 
of carbon atoms N$_\mathrm{C}$. The lowest panel shows the 
absolute total integrated IR absorption cross\textendash sections $I$ (in km/mol), 
divided by N$_\mathrm{C}$.}
\label{total}
\end{figure}

\begin{figure}
\includegraphics[width=\hsize]{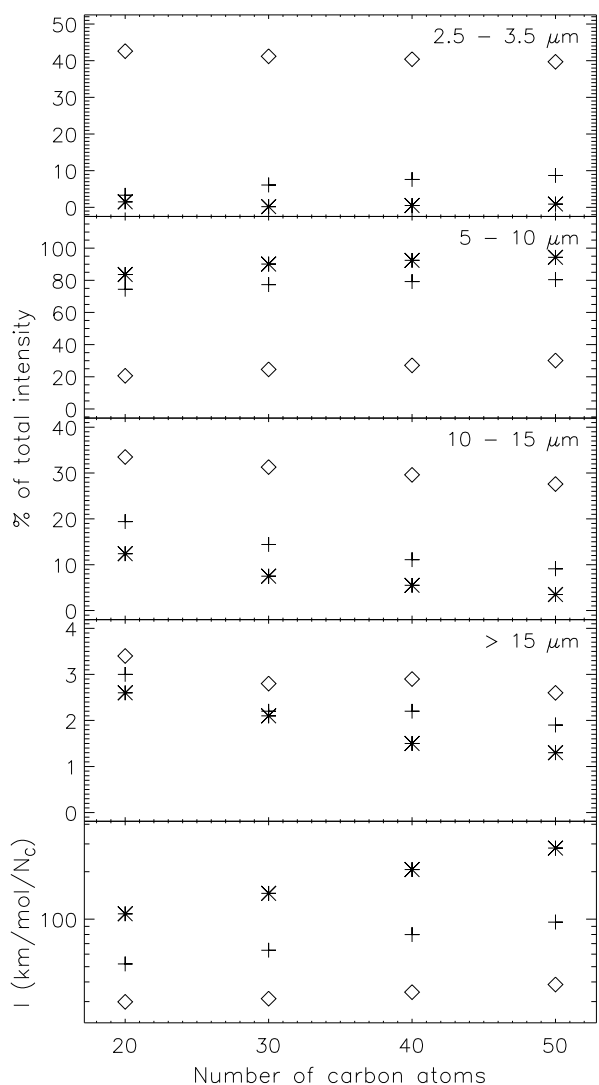}
\caption{Same as Fig.~\ref{total}, restricted to the PAHs of the 
oligoarylenes series, i.e., perylene, terrylene, quaterrylene and
pentarylene (C$_{10n}$H$_{4n+4}$, with n=2, 3, 4, 5).}
\label{rylenes}
\end{figure}

All the computed harmonic vibrational frequencies of the PAHs$^{++}$ considered,
as well for their corresponding anions, neutrals and cations, can be found in 
our on\textendash line database of computed spectral properties of PAHs, 
which is presently under construction \citep{mal06a}. We report here only 
comparative plots for the IR properties of the PAHs under study in their 0, 
+1 and +2 charge states. In agreement with previous works 
\citep{ell99,bau00,pau02}, we found that 
while the harmonic vibrational frequencies are only slightly affected by 
single and double ionisation, absolute and relative IR absorption 
cross\textendash sections undergo significant variations with charge. Figure 
\ref{total} displays the fractions of the total integrated IR absorption 
cross\textendash section in the four spectral ranges $2.5\textendash3.5$, $5\textendash10$, $10\textendash15$~$\mu$m and
$>15$~$\mu$m as a function of the total carbon atoms in the molecule. These 
wavelength intervals are somewhat arbitrary but roughly contain the spectral 
signatures of, respectively: C-H stretchings ($2.5\textendash3.5~\mu$m); C-C in-plane 
stretchings and C-H in-plane bendings ($5\textendash10~\mu$m); C-H out of plane 
bendings ($10\textendash15$~$\mu$m); in-plane and out-of-plane cycle deformations of 
the whole carbon skeleton ($>15$~$\mu$m). The lowest panel of Fig.~\ref{total} 
shows the absolute value of the total integrated IR absorption cross\textendash section 
divided by the number of carbon atoms N$_\mathrm{C}$.
In Fig.~\ref{total} we omitted the 
vibrational data we obtained for the more symmetric molecules, for which 
Jahn\textendash Teller distortion is observed upon ionisation. For example, in the 
specific cases of the cations of triphenylene and coronene, two different 
Jahn\textendash Teller distorted structures are 
known to exist \citep{tor99,kat99}. These two configurations are very close in 
energy and the determination of the most stable structure is a delicate 
problem that may complicate DFT\textendash based calculations and severely affect the 
resulting IR spectra \citep{oom01}. We report in Fig.~\ref{rylenes} the 
corresponding data of Fig.~\ref{total} given only for a specific series of 
PAHs, namely the oligorylenes perylene, terrylene, quaterrylene 
and pentarylene (C$_{10n}$H$_{4n+4}$, with n=2, 3, 4, 5).

As to the accuracy of our \mbox{B3LYP/4\textendash31G} data, this method seems to 
overestimate the intensities of C\textendash H stretching modes \citep{lan96,bau97}. 
The evaluation of absolute IR absorption cross\textendash sections is a difficult task 
since they are known to be strongly affected by the specific choice of the 
basis set employed \citep{lan96,bau97}. This holds true, in particular, for 
open\textendash shell species, that may present problems of spin contamination or 
charge localisation artifacts. As an example, the total integrated IR
absorption cross\textendash sections we obtain for pyrene neutral, cation and dication 
are 464.4, 784.6 and 1758.3 km/mol, while the calculations of \citet{ell99} 
performed at the higher \mbox{B3LYP/6\textendash31G$^\star$} level, give 399.3, 720.9 and 
1613.5 km/mol, respectively. The intensity ratios in the same wavelength 
intervals introduced above are almost coincident among the two sets of results 
for C$_{16}$H$_{10}$ and C$_{16}$H$_{10}$$^{++}$. In the case of the cation 
C$_{16}$H$_{10}$$^{+}$, instead, these values amount to 2, 68, 27 and 3~\%, to be 
compared with 1, 71, 19 and 9~\%, as obtained by \citet{ell99}.  

\subsection{Photo\textendash absorption spectra}

As for the computed vibrational properties, all the theoretical 
photo\textendash absorption spectra of the PAHs$^{++}$ considered and their neutral 
and singly\textendash ionised counterparts can be found in our on\textendash line database. 
For the low\textendash lying $\pi^\star\gets\pi$ transitions occurring in the 
near\textendash IR, visible and near\textendash UV spectral ranges, the application of this 
approach to PAHs and PAHs$^{\pm}$ was shown to be as accurate as previously 
published theoretical results, compared to the available experimental data 
\citep{mal04,mal05}. However, while we established the use of the 
\textsc{octopus} code to calculate reasonably accurate photo\textendash absorption 
spectra of neutral PAHs up to the vacuum\textendash UV spectral range, experimental 
spectra in the far\textendash UV are needed for a similar direct validation for PAH ions.
The computed vertical excitation energies are expected to be precise to within 
a few tenths of an eV, which are the typical accuracies of TD\textendash DFT calculations
employing the currently available exchange\textendash correlation functionals 
\citep{hir99,hir03,wei01,wei03,wei05}. More specifically, depending on the 
character of the specific transition, TD\textendash DFT can reach the same accuracy of 
more sophisticated and expensive wave\textendash function based methods or show 
substantial errors, as in the case of the lowest short\textendash axis polarized states 
in the oligoacenes \citep{gri03}. 

\begin{figure}
\includegraphics[width=\hsize]{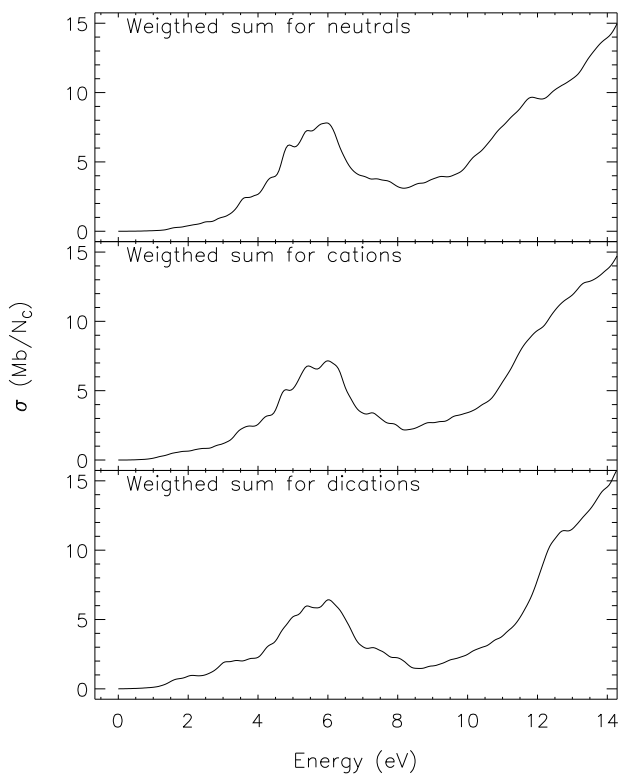}
\caption{Comparison between the weighted sum of the computed
photo\textendash absorption cross\textendash sections $\sigma(E)$ (expressed in 
Megabarns/$\mathrm{N}_\mathrm{C}$, 1~Mb = 10$^{-18}$~cm$^2$) for PAH neutrals, 
the corresponding cations and dications. We used as weight the inverse of the 
total number of carbon atoms $\mathrm{N}_\mathrm{C}$ of each molecule.}
\label{weighted}

\end{figure}

\begin{figure}
\includegraphics[width=\hsize]{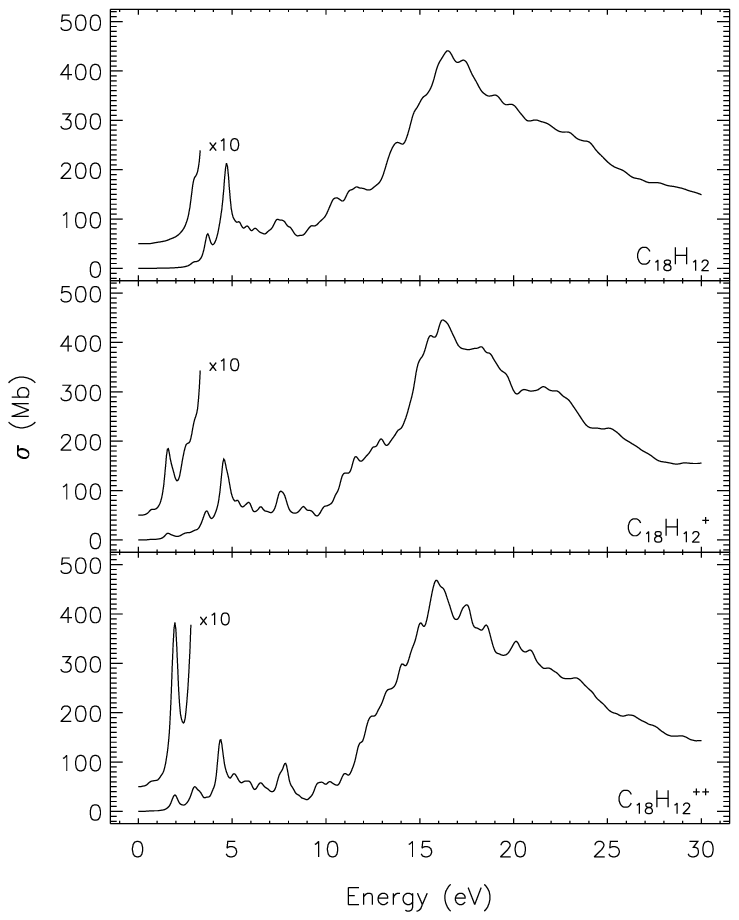}
\caption{Photo\textendash absorption cross\textendash section  $\sigma(E)$ of benzo[a]anthracene
(C$_{18}$H$_{12}$) in the three charge states 0, +1 and +2. Units for
 $\sigma(E)$ are Megabarns (1~Mb~=~10$^{-18}$~cm$^2$).}
\label{benzoanthracene}
\end{figure}

\begin{figure}
\includegraphics[width=\hsize]{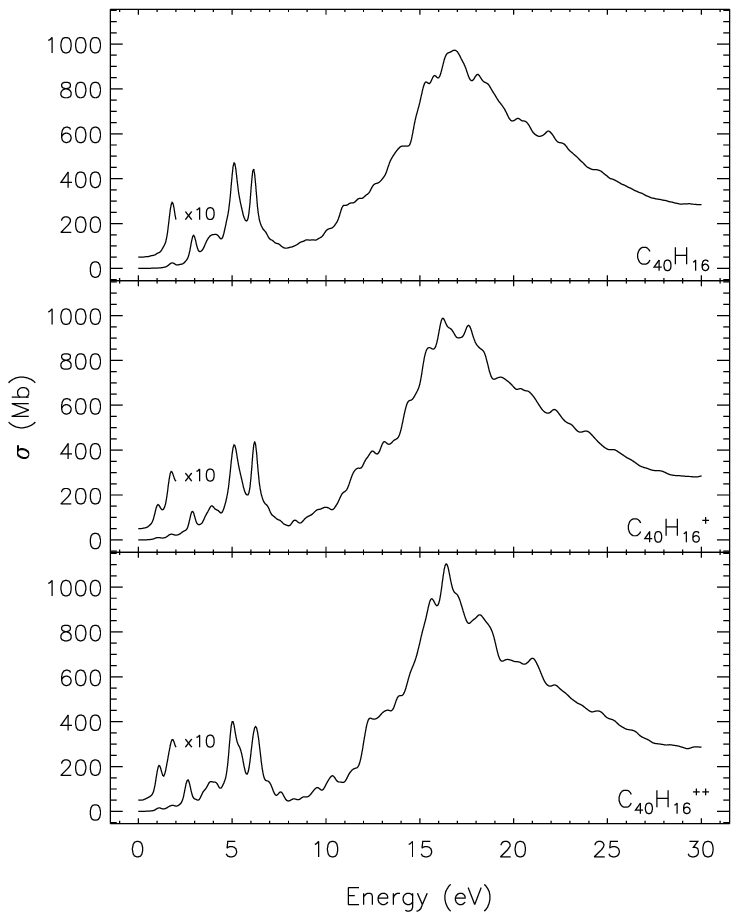}
\caption{Same as Fig.~\ref{benzoanthracene} for circumanthracene 
(C$_{40}$H$_{16}$).}
\label{circumanthracene}
\end{figure}

The spectra obtained for PAHs$^{++}$ can be compared with the corresponding ones 
of PAHs and PAHs$^{+}$. Fig.~\ref{weighted} shows weighted averages of the 
electronic photo\textendash absorption spectra of the same sample of PAHs, respectively 
in the neutral, cationic and dicationic state. To make apparent the changes 
occurring for specific molecules, Figs.~\ref{benzoanthracene} and 
\ref{circumanthracene} compare the photo\textendash absorption cross\textendash sections of
benzo[a]anthracene (C$_{18}$H$_{12}$) and circumanthracene (C$_{40}$H$_{16}$) as a
function of increasing positive charge state. 

\section{Discussion} \label{discussion}

Concerning the computed ionisation energies, the values we found for the 
second AIEs confirm that PAHs$^{++}$ can be expected to be produced by 
two\textendash step ionisation in the ISM. 

As to the IR properties, the positions of the bands are relatively 
similar for the same PAH in different charge states, as already noted 
in previous studies on a few PAH dications \citep{ell99,bau00,pau02}. 
We confirm that there are systematic shifts in some band positions as
a function of ionisation state, which only slightly change the general
pattern. We here concentrate on the much larger
variations in the integrated absorption cross\textendash sections. The latter change, 
for some bands, by one order of magnitude, due to drastic changes in charge 
distribution following ionisation. Our sample of molecules is quite
heterogeneous and Fig.~\ref{total} may be somewhat confusing due to crowding. 
However, general trends versus molecular size and versus charge can still be 
observed, and are more evident upon examination of sequences of molecules of 
the same class, as Fig.~\ref{rylenes} for the oligorylenes series clearly 
shows. 
From the top panels of Figs.~\ref{total} and \ref{rylenes} we see the 
well\textendash known collapse of the intensities of 
C\textendash H stretching modes for singly\textendash ionised species \citep{def93,lan96}. The same 
is observed for doubly\textendash ionised species \citep{bau00} with the exception of 
small PAHs up to $\sim$16 carbon atoms, in which the intensity of these modes 
is partly restored by double ionisation \citep{ell99,pau02}.
A net increase in the intensity of the in-plane C\textendash H band and C\textendash C stretching 
vibrations is observed for PAHs$^+$ with respect to neutral species in the 
$5-10$~$\mu$m window. This occurs even to a larger extent for PAHs$^{++}$, 
whose absorption spectra tend to be dominated only by the features in this 
region. The opposite is observed in the $10-15$~$\mu$m interval, where the 
out-of-plane C\textendash H vibrations are the second major component for neutral 
species but tend to be less important for PAHs$^+$ and PAHs$^{++}$. Bands at
wavelengths larger than $15~\mu$m contribute for less than 10~\% of the 
total integrated IR absorption cross\textendash section $I$. The absolute value of the 
latter quantity, for all of the molecules considered, shows a marked increase 
upon ionisation, as clearly demonstrated by the lowest panel of 
Figs.~\ref{total} and \ref{rylenes}. On the average we found that PAHs$^{++}$ 
absorb about $2.5\pm1.0$ times more than their parent singly ionised molecules 
and about $4.7\pm1.1$ times more than their parent neutrals.

\begin{figure}
\includegraphics[width=\hsize]{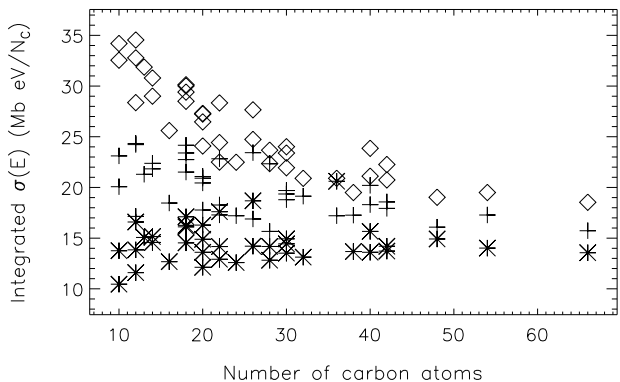}
\caption{Comparison between the integrated values between $\sim$7 and $\sim$12~eV 
of the photo\textendash absorption cross\textendash sections $\sigma(E)$ divided by the number of 
carbon atoms N$_\mathrm{C}$, for the neutrals (diamonds), cations (crosses) and 
dications (asterisks) of the PAHs considered, as a function of N$_\mathrm{C}$.}
\label{gap}
\end{figure}

With respect to the electronic absorption properties, inspection of 
Figs.~\ref{benzoanthracene} and \ref{circumanthracene} shows that all 
the spectra for specific molecules show comparable  features in the UV range, 
displaying the same behaviour already known for neutral species 
\citep{job92a,job92b,lea95b}, i.e., increasing smoothly at energies above 
$\sim7.4$~eV to a maximum at $\sim17.4$~eV in a single broad absorption peak 
composed of \mbox{$\sigma^*\gets\sigma$},  \mbox{$\sigma^*\gets\pi$}, \mbox{$\pi^*\gets\sigma$} and Rydberg 
spectral transitions. The onset of this latter broad absorption moves
blueward and becomes steeper with increasing positive charge, following
the increase in subsequent ionisation potentials. This translates in 
a systematic decrease of the cross\textendash section between $\sim$7 and $\sim$12~eV
with increasing positive charge, as shown in Fig.~\ref{gap}.
In the near\textendash IR, visible and near\textendash UV spectral ranges, 
the dication has strong absorption features of $\pi^\star\gets\pi$ character like the 
singly\textendash ionised species. Such a behaviour is an expected consequence of the 
different one\textendash electron energy levels and of the resulting first few electric 
dipole\textendash permitted transitions. 


More generally, Fig.~\ref{weighted} shows that the three averaged spectra 
present very similar features. The main difference is the progressive decrease,
with increasing positive charge, of the cross\textendash section between $\sim$7 and 
$\sim$12~eV, as expected from the behaviour of the individual spectra in this 
same energy interval (cf.~Fig.~\ref{gap}). 
Both the former similarity and the latter difference have interesting
astrophysical implications for the UV range of the spectrum: apart from small 
details (i.e., the precise position of single, specific transitions of a 
single, specific molecule), the contribution of any given PAH will depend 
only slightly on its ionisation state. The average over a sample 
of molecules has the main effect of smoothing out the detailed structure 
resulting from the contribution of each single species. This confirms 
our previous findings \citep{job92a,job92b,mal04,mal05}, extending 
them to PAHs$^{++}$: if \emph{any} PAHs of small to medium size, be they 
neutral, cationic or dicationic are to account for the far\textendash UV rise of
the interstellar extinction curve they \emph{must} also contribute to the 
short\textendash wavelength side of the interstellar extinction bump at $\sim$2175~\AA{} 
($\sim$5.7~eV). On the other hand, the onset of the far\textendash UV rise produced by
PAHs ought to depend on their charge state, therefore comparison with
observed extinction curves could provide an observational handle to estimate
the average charge state of interstellar PAHs.

The photo\textendash stability of PAHs$^{++}$ deserves further investigation: given 
the photo\textendash absorption spectra we obtained, PAHs$^{++}$ will efficiently 
absorb UV photons, if available, and can be expected to convert at least 
a fraction of the resulting energy to internal excitation, possibly 
reaching their dissociation threshold. Should this be the case, 
this might turn out to be a major destruction channel for PAHs. However, 
PAHs$^{++}$ absorb slightly less than their parent neutrals and singly ionised 
species between $\sim$7 and 
$\sim$12~eV, and cool faster by IR fluorescence since, as we saw before, 
their vibrational transitions are overall from two to five times more 
intense. Since in these molecules unimolecular dissociation is 
expected to occur in competition with vibrational cooling, PAHs$^{++}$ should
actually be much more stable against photodissociation than PAHs and 
PAHs$^{+}$ \emph{if} the dissociation thresholds are nearly unchanged by 
ionisation. More laboratory studies on the relaxation paths of PAHs$^{++}$
following photon absorption are needed.

\section{Conclusions and future work}
\label{conclusions}


We performed a systematic theoretical study on the ionisation energies of PAHs 
and the IR and visible\textendash UV absorption properties of their dications. The values
we found for the second IEs confirm that PAHs could reach the doubly\textendash ionised 
state in H$_\mathrm{I}$ regions. Combining the IR and visible\textendash UV absorption 
properties of PAHs$^{++}$ we predict an increased photostability for these 
species with respect to their neutral and singly\textendash ionized parent molecules,
unless they have lowest dissociation channels.

The recent paper by \citet{wit06}
has recently renewed the interest in the dications of PAHs, proposing them as
plausible candidates of the Extended Red Emission. The fundamental data we 
present here for a large sample of different PAHs will be useful to model in 
detail the photophysics of these molecules in interstellar conditions.
For instance, we can estimate quantitatively the complete IR emission spectrum 
of each molecule in the sample as a function of its charge state, for direct 
comparison with data from the Infrared Space Observatory and Spitzer Space 
Telescope missions and, in the near future, from the Herschel Space 
Observatory mission \citep{mul06a,mul06b,mul06c}. In addition, combined with 
measurements of the electron recombination rates of singly\textendash ionized PAHs 
\citep{nov05}, and their extension to doubly\textendash ionized species, which still 
have to be studied, the present work opens the way for a detailed modelling 
of the chemistry and photophysics of PAHs in interstellar environments 
\citep{bak98,lep01} where they may play a prominent role.

\begin{acknowledgements}
G.~Malloci acknowledges the financial support by the 
``Minist\`ere de la Recherche''. We warmly thank the authors of \textsc{octopus} for making their code 
available under a free license. We acknowledge the High Performance 
Computational Chemistry Group for the use of NWChem, A Computational 
Chemistry Package for Parallel Computers, Version~4.7 (2005), PNNL, Richland, 
Washington, USA. Part of the calculations were performed using CINECA and 
CALMIP supercomputing facilities. 

\end{acknowledgements}


\appendix
\section{Computed adiabatic and vertical single and double IEs}

The following Table.~\ref{double} lists the single and double adiabatic and
vertical IEs of each molecule considered in this work. As explained in the
body of the paper, the single and double AIEs were computed, respectively, 
via total energy differences between the geometry optimised cations and 
dications and the total energy of the geometry optimised neutral. The 
corresponding VIEs were evaluated as the differences between the single point 
energy of the cation and dication at the optimised neutral geometry and the 
total energy of the optimised neutral. The adiabatic second ionisation energy 
$\Delta I$ is simply given by $I^{++}_\mathrm{ad} \textendash{} I^+_\mathrm{ad}$ while its vertical 
value is obtained through the difference between the single point energy of 
the dication at the optimised cation geometry and the total energy of the 
optimised cation. We also evaluated the energy  difference between the direct 
vertical double ionisation process (PAH$\to$PAH$^{++}$) and the intermediate one 
in which the mono\textendash cation is allowed to relax to its ground\textendash state geometry 
before second ionisation (PAH$\to$PAH$^+$$\to$PAH$^{++}$). The last two columns of 
Table~\ref{double} report the sum of the first and second vertical ionisation 
energies $I^+_\mathrm{V} + \Delta I_\mathrm{V}$ and the difference between the double 
vertical ionisation energy  $I^{++}_\mathrm{V}$ and $I^+_\mathrm{V} + \Delta I_\mathrm{V}$. 

For all singlet dications under study it is found that the intermediate 
formation of the monocation is less energy demanding than the direct double 
ionisation process (the above difference being positive); the opposite sign is 
instead obtained for almost all triplet dications. As pointed out by 
\citet{sch01}, this comes from the triplet dication geometries being more 
similar to the neutral ground\textendash state geometries than to the ones of the 
monocation.

\onecolumn

\begin{landscape}

{\footnotesize
\begin{longtable}{cc ccc ccc ccc|cc}
\caption{Adiabatic and vertical ionisation energies (in eV) of each PAH in 
our sample evaluated at the \mbox{B3LYP/6-31+G$^\star$} level of theory. 
Whenever
available we list laboratory data for comparison. The experimental single IEs 
come from the NIST Chemistry WebBook \citep{lia05} while second and double IEs 
are, respectively, from charge stripping and photon impact measurements of 
\citet{sch01} for corannulene and coronene and from \citet{tob94} in all the 
other cases. According to our calculations the dication electronic 
ground\textendash states are marked in boldface.} 
\label{double}\\
\hline \hline
\noalign{\smallskip}
PAH & \multicolumn{3}{c}{First IE ($I^+$)} & Dication & 
\multicolumn{3}{c}{Second IE ($\Delta I$)} & \multicolumn{3}{c|}
{Double IE ($I^{++}$)}&&\\
molecule & Adiab. & Vert. & Exp. & state & Adiab. & 
Vert. & Exp. & Adiab. & Vert. & Exp. & $I^+_\mathrm{V} 
+ \Delta I_\mathrm{V}$ & $I^{++}_\mathrm{V}$ \textendash{} ($I^+_\mathrm{V} + \Delta I_\mathrm{V}$)\\
\noalign{\smallskip}
\hline
\noalign{\smallskip}
\endfirsthead
\multicolumn{13}{c}{{\bfseries \tablename\ \thetable{}} 
-- continued from previous page} \\ \hline \hline
\noalign{\smallskip}
PAH & \multicolumn{3}{c}{First IE ($I^+$)} & Dication & 
\multicolumn{3}{c}{Second IE ($\Delta I$)} & 
\multicolumn{3}{c|}{Double IE ($I^{++}$)}&&\\
molecule & Adiab. & Vert. & Exp. & state & Adiab. & 
Vert. & Exp. & Adiab. & Vert. & Exp. & $I^+_\mathrm{V} 
+ \Delta I_\mathrm{V}$ & $I^{++}_\mathrm{V}$ \textendash{} ($I^+_\mathrm{V} + \Delta I_\mathrm{V}$)\\
\noalign{\smallskip}
\noalign{\smallskip}
\endhead
\multicolumn{13}{c}{{continued on next page}} \\ \hline \endfoot
\hline 
\noalign{\smallskip}
\endlastfoot
Azulene & 
\multirow{2}*{7.15} & \multirow{2}*{7.27} & \multirow{2}*{7.42$\pm$0.02} & 
{\bf singlet} & {\bf 12.77} & {\bf 12.87} & \multirow{2}*{\textemdash} & 
{\bf 19.92} & {\bf 20.34} & \multirow{2}*{\textemdash} & {\bf 20.14} & {\bf + 0.20} \\
(C$_{10}$H$_{8}$) & & & & triplet & 13.39 & 13.53 & & 20.54 & 20.63 & & 20.80 
& \textendash{} 0.17 \\
\noalign{\smallskip}
\hline
\noalign{\smallskip}
Naphthalene &  
\multirow{2}*{7.80} & \multirow{2}*{7.89} & 
\multirow{2}*{8.144$\pm$0.001} & {\bf singlet} & {\bf 13.19} & {\bf 13.28} & 
\multirow{2}*{14.6$\pm$0.5} & {\bf 20.99} & {\bf 21.35} & 
\multirow{2}*{21.5$\pm$0.2} & {\bf 21.17} & {\bf + 0.18}\\
(C$_{10}$H$_{8}$) & & & & triplet & 13.65 & 13.77 & & 21.45 & 21.57 & & 21.66 
& \textendash{} 0.09\\
\noalign{\smallskip}
\hline
\noalign{\smallskip}
Acenaphtylene &  
\multirow{2}*{7.66} & \multirow{2}*{7.86} & 
\multirow{2}*{8.12$\pm$0.10} & singlet & 12.81 & 13.06 & \multirow{2}*{\textemdash} 
& 20.48 & 21.32 & \multirow{2}*{21.4$\pm$0.2} & 20.92 & + 0.39\\
(C$_{12}$H$_{8}$) & & & & {\bf triplet} & {\bf 12.78} & {\bf 12.88} & & 
{\bf 20.45} & {\bf 20.66} & & {\bf 20.74} & {\bf \textendash{} 0.08} \\
\noalign{\smallskip}
\hline
\noalign{\smallskip}
Biphenylene &  
\multirow{2}*{7.22} & \multirow{2}*{7.37} & 
\multirow{2}*{7.58$\pm$0.03} & {\bf singlet} & {\bf 12.09} & {\bf 
12.26} & 
\multirow{2}*{\textemdash} & {\bf 19.31} & {\bf 19.94} & 
\multirow{2}*{19.7$\pm$0.2} & 
{\bf 19.63} & {\bf + 0.31}\\
(C$_{12}$H$_{8}$)& & & & triplet & 13.23 & 13.31 & & 20.45 & 20.61 & & 
20.68 & \textendash{} 0.08 \\
\noalign{\smallskip}
\hline
\noalign{\smallskip}
Acenaphthene &  
\multirow{2}*{7.37} & \multirow{2}*{7.46} & 
\multirow{2}*{7.75$\pm$0.05} & {\bf singlet} & {\bf 12.42} & {\bf 12.51}& 
\multirow{2}*{\textemdash}& {\bf 19.78} & {\bf 20.17} & \multirow{2}*{21.7$\pm$0.2} & 
{\bf 19.97}& {\bf + 0.20}\\
(C$_{12}$H$_{10}$) & & & & triplet & 13.22 & 13.35 & & 20.59 & 20.72 & & 20.81 
& \textendash{} 0.09 \\
\noalign{\smallskip}
\hline
\noalign{\smallskip}
Fluorene &  
\multirow{2}*{7.56} & \multirow{2}*{7.69} & 
\multirow{2}*{7.91$\pm$0.02} & {\bf singlet} & {\bf 12.39} & {\bf 12.52} & 
\multirow{2}*{\textemdash} & {\bf 19.96} & {\bf 20.47} & \multirow{2}*{21.0$\pm$0.2} & 
{\bf 20.22}& {\bf + 0.26}\\ 
(C$_{13}$H$_{10}$) & & & & triplet & 12.86 & 12.99 & & 20.43 & 20.59 & & 20.68 
& \textendash{} 0.09 \\
\noalign{\smallskip}
\hline
\noalign{\smallskip}
Anthracene &  
\multirow{2}*{7.02} & \multirow{2}*{7.09} & 
\multirow{2}*{7.439$\pm$0.006} & {\bf singlet} & {\bf 11.68} & {\bf 11.74} & 
\multirow{2}*{\textemdash} & {\bf 18.70} & {\bf 18.95} & \multirow{2}*{\textemdash} & 
{\bf 18.83} & {\bf + 0.12}\\ 
(C$_{14}$H$_{10}$) & & & & triplet & 12.68 & 12.77 & & 19.70 & 19.80 & & 19.86 &
\textendash{} 0.06\\
\noalign{\smallskip}
\hline
\noalign{\smallskip}
Phenanthrene &  
\multirow{2}*{7.53} & \multirow{2}*{7.63} & 
\multirow{2}*{7.891$\pm$0.001} & {\bf singlet} & {\bf 12.11} & {\bf 12.23}
& \multirow{2}*{\textemdash} & {\bf 19.63} & {\bf 20.08} & \multirow{2}*{\textemdash} & 
{\bf 19.86}& {\bf + 0.22}\\ 
(C$_{14}$H$_{10}$) & & & & triplet & 12.19 & 12.32 & & 19.72 & 19.86 & & 19.96 &
\textendash{} 0.09 \\
\noalign{\smallskip}
\hline
\noalign{\smallskip}
Pyrene &  
\multirow{2}*{7.07} & \multirow{2}*{7.14} & 
\multirow{2}*{7.426$\pm$0.001} & {\bf singlet} & {\bf 11.57} & {\bf 11.64} & 
\multirow{2}*{14.4$\pm$0.5} & {\bf 18.63} & {\bf 18.92} & 
\multirow{2}*{19.8$\pm$0.2} & {\bf 18.78} & {\bf + 0.14}\\ 
(C$_{16}$H$_{10}$) & & & & triplet & 12.28 & 12.39 & & 19.35 & 19.44 & & 19.53 &
\textendash{} 0.09 \\
\noalign{\smallskip}
\hline
\noalign{\smallskip}
Tetracene &  
\multirow{2}*{6.49} & \multirow{2}*{6.55} & 
\multirow{2}*{6.97$\pm$0.05} & {\bf singlet} & {\bf 10.66} & {\bf 10.70} & 
\multirow{2}*{13.1$\pm$0.5} & {\bf 17.15} & {\bf 17.34} & 
\multirow{2}*{18.6$\pm$0.2} & {\bf 17.25} & {\bf + 0.09}\\ 
(C$_{18}$H$_{12}$) & & & & triplet & 11.47 & 11.55 & & 17.96 & 18.11 & & 18.10 &
+ 0.01 \\
\noalign{\smallskip}
\hline
\noalign{\smallskip}
Chrysene &  
\multirow{2}*{7.17} & \multirow{2}*{7.25} & 
\multirow{2}*{7.60$\pm$0.01} & {\bf singlet} & {\bf 11.28} & {\bf 11.37} & 
\multirow{2}*{\textemdash} & {\bf 18.45} & {\bf 18.78} & \multirow{2}*{\textemdash} & 
{\bf 18.63} & {\bf + 0.16}\\ 
(C$_{18}$H$_{12}$) & & & & triplet & 11.67 & 11.75 & & 18.84 & 18.94 & & 19.01 &
\textendash{} 0.06\\
\noalign{\smallskip}
\hline
\noalign{\smallskip}
Benz[a]anthracene &  \multirow{2}*{6.98} & 
\multirow{2}*{7.05} & \multirow{2}*{7.45$\pm$ 0.05} & {\bf singlet} & 
{\bf 11.16} & {\bf 11.24} & \multirow{2}*{\textemdash} & {\bf 18.14} & {\bf 18.42} & 
\multirow{2}*{\textemdash} & {\bf 18.29} & {\bf + 0.13}\\ 
(C$_{18}$H$_{12}$) & & & & triplet & 11.35 & 11.45 & & 18.33 & 18.48 & & 
18.50 & \textendash{} 0.02\\
\noalign{\smallskip}
\hline
\noalign{\smallskip}
Triphenylene &  
\multirow{2}*{7.52} & \multirow{2}*{7.86} & 
\multirow{2}*{7.87$\pm$0.02} & singlet & 11.72 & 11.81 & \multirow{2}*{\textemdash} & 
19.24 & 19.59 & \multirow{2}*{\textemdash} & 19.67 & \textendash{} 0.08\\ 
(C$_{18}$H$_{12}$) & & & & {\bf triplet} & {\bf 11.57} & {\bf 11.65} & & 
{\bf 19.09} & {\bf 19.20} & & {\bf 19.51} & {\bf \textendash{} 0.31} \\
\noalign{\smallskip}
\hline
\noalign{\smallskip}
Corannulene &  
\multirow{2}*{7.62} & \multirow{2}*{7.73} & \multirow{2}*{7.83$\pm$0.02} & 
singlet & 12.02 & 12.15 & \multirow{2}*{13.8$\pm$0.3}
& 19.63 & 20.14 & \multirow{2}*{20.1$\pm$0.2} & 19.88 & + 0.26 \\ 
(C$_{20}$H$_{10}$) & & & & {\bf triplet} & {\bf 11.66} & {\bf 11.76} & 
& {\bf 19.28} & {\bf 19.50} & & {\bf 19.49} & {\bf + 0.01} \\
\noalign{\smallskip}
\hline
\noalign{\smallskip}
Benzo[a]pyrene &  \multirow{2}*{6.71} & 
\multirow{2}*{6.78} & \multirow{2}*{7.12$\pm$0.01} & {\bf singlet} & 
{\bf 10.75} & {\bf 10.82} & \multirow{2}*{\textemdash} & {\bf 17.46} & 
{\bf 17.73} & \multirow{2}*{\textemdash} & {\bf 17.60} & {\bf +0.13}\\ 
(C$_{20}$H$_{12}$) & & & & triplet & 11.47 & 11.58 & & 18.18 & 18.30 & 
& 18.36 & \textendash0.06\\
\noalign{\smallskip}
\hline
\noalign{\smallskip}
Benzo[e]pyrene &  
\multirow{2}*{7.05} & \multirow{2}*{7.12} & 
\multirow{2}*{7.43$\pm$0.04} & {\bf singlet} & {\bf 11.20} & {\bf 11.26} & 
\multirow{2}*{\textemdash} & {\bf 18.25} & {\bf 18.51} & 
\multirow{2}*{\textemdash} & {\bf 18.38} & {\bf + 0.13}\\ 
(C$_{20}$H$_{12}$) & & & & triplet & 11.34 & 11.42 & & 18.38 & 18.52 & 
& 18.54 & \textendash0.02\\
\noalign{\smallskip}
\hline
\noalign{\smallskip}
Perylene &  
\multirow{2}*{6.57} & \multirow{2}*{6.64} & 
\multirow{2}*{6.960$\pm$0.001} & {\bf singlet} & {\bf 10.62} & {\bf 
10.68} & \multirow{2}*{\textemdash} & {\bf 17.19} & {\bf 17.46} & 
\multirow{2}*{18.0$\pm$0.2} & {\bf 17.33} & {\bf + 0.14}\\ 
(C$_{20}$H$_{12}$) & & & & triplet & 12.03 & 12.11 & & 18.60 & 18.73 & & 18.75 &
\textendash0.02\\
\noalign{\smallskip}
\hline
\noalign{\smallskip}
Anthanthrene &  
\multirow{2}*{6.46} & \multirow{2}*{6.52} & 
\multirow{2}*{6.92$\pm$0.04} & {\bf singlet} & {\bf 10.43} & {\bf 10.48} & 
\multirow{2}*{\textemdash} & {\bf 16.90} & {\bf 17.10} & \multirow{2}*{\textemdash} & 
{\bf 17.00} & {\bf + 0.10} \\ 
(C$_{22}$H$_{12}$) & & & & triplet & 11.38 & 11.54 & & 18.01 & 18.09 & & 
18.06 & +0.03 \\
\noalign{\smallskip}
\hline
\noalign{\smallskip}
Benzo[g,h,i]perylene &  \multirow{2}*{6.79} & 
\multirow{2}*{6.86} & \multirow{2}*{7.17$\pm$0.02} & {\bf singlet} & 
{\bf 10.74} & {\bf 10.80} & \multirow{2}*{\textemdash} 
& {\bf 17.53} & {\bf 17.78} & \multirow{2}*{\textemdash} & {\bf 17.65} & {\bf + 0.13}\\
(C$_{22}$H$_{12}$) & & & & triplet & 11.19 & 11.28 & & 17.98 & 18.09 & & 18.14
& \textendash{} 0.05\\
\noalign{\smallskip}
\hline
\noalign{\smallskip}
Pentacene &  
\multirow{2}*{6.12} & \multirow{2}*{6.16} & 
\multirow{2}*{6.63$\pm$0.05} & {\bf singlet} & {\bf 9.91} & {\bf 9.94} & 
\multirow{2}*{12.8$\pm$0.5} & {\bf 16.03} & {\bf 16.18} & 
\multirow{2}*{17.4$\pm$0.2} & {\bf 16.11} & {\bf + 0.07} \\ 
(C$_{22}$H$_{14}$) & & & & triplet & 10.56 & 10.63 & & 16.67 & 16.79 & & 16.79
& 0.00 \\
\noalign{\smallskip}
\hline
\noalign{\smallskip}
Coronene &  \multirow{2}*{7.02} & 
\multirow{2}*{7.08} & \multirow{2}*{7.29$\pm$0.03} & singlet & 10.87 & 
10.92 & \multirow{2}*{12.8$\pm$0.3} & 17.88 & 18.11 & 
\multirow{2}*{18.5$\pm$0.2} & 18.00 & + 0.11 \\ 
(C$_{24}$H$_{12}$) & & & & {\bf triplet} & {\bf 10.72} & {\bf 10.79} & 
& {\bf 17.74} & {\bf 17.80} & & {\bf 17.87} & {\bf \textendash0.06}\\
\noalign{\smallskip}
\hline
\noalign{\smallskip}
Dibenzo[b,def]crysene &  
\multirow{2}*{6.39} & \multirow{2}*{6.45} & 
\multirow{2}*{6.82} & {\bf singlet} & {\bf 10.11} & {\bf 10.17} & 
 \multirow{2}*{\textemdash} & {\bf 16.50} & {\bf 16.73} & \multirow{2}*{\textemdash}& 
{\bf 16.62} & {\bf + 0.11} \\ 
(C$_{24}$H$_{14}$) & & & & triplet & 11.00 & 11.06 & & 17.39 & 17.51 & & 17.51
& 0.00 \\
\noalign{\smallskip}
\hline
\noalign{\smallskip}
Dibenzo[cd,lm]perylene &  
\multirow{2}*{6.31} & \multirow{2}*{6.38} & 
\multirow{2}*{6.72$\pm$0.02} & {\bf singlet}& {\bf 9.88} & {\bf 9.94} & 
\multirow{2}*{\textemdash} & {\bf 16.20} & {\bf 16.45} & \multirow{2}*{\textemdash} & 
{\bf 16.32} & {\bf + 0.13} \\ 
(C$_{26}$H$_{14}$) & & & & triplet & 11.03 & 11.12 & & 17.35 & 17.43 & & 
17.50 & \textendash{} 0.07\\
\noalign{\smallskip}
\hline
\noalign{\smallskip}
Hexacene &  
\multirow{2}*{5.83} & \multirow{2}*{5.87} & \multirow{2}*{6.36$\pm$0.02} & {\bf singlet} & {\bf 9.34} & {\bf 9.37} & \multirow{2}*{\textemdash}  
& {\bf 15.18} & {\bf 15.30} & \multirow{2}*{\textemdash} & {\bf 15.24} & {\bf +0.06}\\
(C$_{26}$H$_{16}$) & & & & triplet & 9.84 & 9.90 & & 15.68 & 15.78 & & 
15.78 & 0.00 \\
\noalign{\smallskip}
\hline
\noalign{\smallskip}
Bisanthene &  
\multirow{2}*{5.85} & \multirow{2}*{5.90} & \multirow{2}*{6.30} & 
{\bf singlet} & {\bf 9.47} & {\bf 9.51} & \multirow{2}*{\textemdash}  
& {\bf 15.33} & {\bf 15.48} & \multirow{2}*{\textemdash} & {\bf 15.40} & {\bf + 0.08}\\
(C$_{28}$H$_{14}$) & & & & triplet & 10.69 & 10.76 & & 16.54 & 16.70 & & 16.66 &
+ 0.04 \\
\noalign{\smallskip}
\hline
\noalign{\smallskip}
Benzo[a]coronene &  \multirow{2}*{6.75} & 
\multirow{2}*{6.81} & \multirow{2}*{7.08} & 
{\bf singlet} & {\bf 10.34} & {\bf 10.39} & \multirow{2}*{\textemdash}  
& {\bf 17.10} & {\bf 17.30} & \multirow{2}*{\textemdash} & {\bf 17.20} & {\bf + 0.10}\\
(C$_{28}$H$_{14}$) & & & & triplet & 10.46 & 10.52 & & 17.21 & 17.29 & & 17.33
& \textendash{} 0.04\\
\noalign{\smallskip}
\hline
\noalign{\smallskip}
Dibenzo[bc,kl]coronene &  
\multirow{2}*{6.01} & \multirow{2}*{6.05} & 
\multirow{2}*{6.42$\pm$0.02} & {\bf singlet} & {\bf 9.54} & {\bf 9.57} & 
\multirow{2}*{\textemdash} & {\bf 15.55} & {\bf 15.70} & 
\multirow{2}*{\textemdash} & {\bf 15.62} & {\bf + 0.08} \\ 
(C$_{30}$H$_{14}$) & & & & triplet & 10.19 & 10.27 & & 16.20 & 16.36 & 
& 16.32 & + 0.04 \\
\noalign{\smallskip}
\hline
\noalign{\smallskip}
Dibenzo[bc,ef]coronene &  
\multirow{2}*{6.12} & \multirow{2}*{6.17} & 
\multirow{2}*{6.50} & {\bf singlet} & {\bf 9.66} & {\bf 9.69} & 
\multirow{2}*{\textemdash} & {\bf 15.78} & {\bf 15.94} & \multirow{2}*{\textemdash} & 
{\bf 15.86} & {\bf + 0.08} \\ 
(C$_{30}$H$_{14}$) & & & & triplet & 10.67 & 10.78 & & 16.79 & 16.87 & & 16.95 
& \textendash0.08 \\
\noalign{\smallskip}
\hline
\noalign{\smallskip}
Terrylene &  
\multirow{2}*{5.98} & \multirow{2}*{6.05} & 
\multirow{2}*{6.42$\pm$0.02} & {\bf singlet} & {\bf 9.33} & {\bf 9.38} & 
\multirow{2}*{\textemdash} & {\bf 15.31} & {\bf 15.55} & \multirow{2}*{\textemdash} & 
{\bf 15.43} & {\bf + 0.12} \\ 
(C$_{30}$H$_{16}$) & & & & triplet & 10.40 & 10.45 & & 16.38 & 16.49 & & 16.50 
& \textendash{} 0.01 \\
\noalign{\smallskip}
\hline
\noalign{\smallskip}
Ovalene &  
\multirow{2}*{6.36} & \multirow{2}*{6.41} & 
\multirow{2}*{6.71} & {\bf singlet} & {\bf 9.82} & {\bf 9.86} & 
\multirow{2}*{\textemdash} & {\bf 16.18} & {\bf 16.34} & 
\multirow{2}*{\textemdash} & {\bf 16.26} & {\bf + 0.08}\\ 
(C$_{32}$H$_{14}$) & & & & triplet & 10.30 & 10.35 & & 16.66 & 16.72 & 
& 16.76 & \textendash0.04 \\
\noalign{\smallskip}
\hline
\noalign{\smallskip}
Tetrabenzo[bc,ef,kl,no]coronene &  
\multirow{2}*{5.41} & \multirow{2}*{5.44} & \multirow{2}*{\textemdash} & 
{\bf singlet} & {\bf 8.75} & {\bf 8.77} & \multirow{2}*{\textemdash}  
& {\bf 14.16} & {\bf 14.24} & \multirow{2}*{\textemdash} & 
{\bf 14.21} & {\bf + 0.03} \\ 
(C$_{36}$H$_{16}$) & & & & triplet & 9.66 & 9.72 & & 15.07 & 15.17 & & 
15.15 & + 0.02\\
\noalign{\smallskip}
\hline
\noalign{\smallskip}
Circumbiphenyl &  \multirow{2}*{6.46} & 
\multirow{2}*{6.52} & \multirow{2}*{6.81$\pm$0.02} & {\bf singlet} & 
{\bf 9.58} & {\bf 9.64} & \multirow{2}*{\textemdash}  
& {\bf 16.04} & {\bf 16.26} & \multirow{2}*{\textemdash} & {\bf 16.15} & 
{\bf + 0.11} \\ 
(C$_{38}$H$_{16}$) & & & & triplet & 9.70 & 9.75 & & 16.16 & 16.21 & & 
16.27 & \textendash0.06\\
\noalign{\smallskip}
\hline
\noalign{\smallskip}
Circumanthracene &  
\multirow{2}*{5.90} & \multirow{2}*{5.94} & \multirow{2}*{\textemdash} & 
{\bf singlet} & {\bf 9.10} & {\bf 9.13} & \multirow{2}*{\textemdash} 
& {\bf 15.00} & {\bf 15.12} & \multirow{2}*{\textemdash} & {\bf 15.06} & 
{\bf + 0.06} \\ 
(C$_{40}$H$_{16}$) & & & & triplet & 9.76 & 9.82 & & 15.66 & 15.95 & 
& 15.75 & + 0.20 \\
\noalign{\smallskip}
\hline
\noalign{\smallskip}
Quaterrylene &  
\multirow{2}*{5.62} & \multirow{2}*{5.68} & 
\multirow{2}*{6.11$\pm$0.02} & {\bf singlet} & {\bf 8.50} & {\bf 8.55} & 
\multirow{2}*{\textemdash} & {\bf 14.12} & {\bf 14.35} & \multirow{2}*{\textemdash} & 
{\bf 14.23} & {\bf + 0.12}\\ 
(C$_{40}$H$_{20}$) & & & & triplet & 9.36 & 9.40 & & 14.97 & 15.08 & & 15.08 
& 0.00\\
\noalign{\smallskip}
\hline
\noalign{\smallskip}
Circumpyrene &  \multirow{2}*{6.06} & \multirow{2}*{6.10} & 
\multirow{2}*{\textemdash} & {\bf singlet} & {\bf 9.13} & {\bf 9.16} & 
\multirow{2}*{\textemdash} & {\bf 15.19} & {\bf 15.35} & \multirow{2}*{\textemdash} & 
{\bf 15.27} &  {\bf + 0.08} \\ 
(C$_{42}$H$_{16}$) & & & & triplet & 9.68 & 9.73 & & 15.75 & 15.80 & & 15.84 & 
\textendash0.04\\
\noalign{\smallskip}
\hline
\noalign{\smallskip}
Hexabenzocoronene &  
\multirow{2}*{6.55} & \multirow{2}*{6.79} & \multirow{2}*{\textemdash} & 
singlet & 9.58 & 9.63 & \multirow{2}*{\textemdash} & 16.14 & 16.33 & 
\multirow{2}*{\textemdash} & 16.42 & \textendash0.09 \\ 
(C$_{42}$H$_{18}$) & & & & {\bf triplet} & {\bf 9.52} & {\bf 9.57} & 
& {\bf 16.07} & {\bf 16.15} & & {\bf 16.36} & {\bf \textendash0.21} \\
\noalign{\smallskip}
\hline
\noalign{\smallskip}
Dicoronylene &  
\multirow{2}*{6.07} & \multirow{2}*{6.13} & 
\multirow{2}*{\textemdash} & {\bf singlet} & {\bf 8.84} & {\bf 8.89} & 
\multirow{2}*{\textemdash} & {\bf 14.91} & {\bf 15.13} & 
\multirow{2}*{\textemdash} & {\bf 15.02} & {\bf + 0.11} \\ 
(C$_{48}$H$_{20}$) & & & & triplet & 9.37 & 9.40 & & 15.44 & 15.50 & 
& 15.53 & \textendash0.03 \\
\noalign{\smallskip}
\hline
\noalign{\smallskip}
Pentarylene &  
\multirow{2}*{5.37} & \multirow{2}*{5.43} & 
\multirow{2}*{\textemdash} & {\bf singlet} & {\bf 7.92} & {\bf 7.97} & 
\multirow{2}*{\textemdash} & {\bf 13.29} & {\bf 13.51} & 
\multirow{2}*{\textemdash} & {\bf 13.39} & {\bf + 0.11}\\ 
(C$_{50}$H$_{24}$) & & & & triplet & 8.64 & 8.68 & & 14.00 & 14.10 & &
14.11 & \textendash{} 0.01\\
\noalign{\smallskip}
\hline
\noalign{\smallskip}
Circumcoronene &  \multirow{2}*{6.14} & \multirow{2}*{6.35} & 
\multirow{2}*{\textemdash} & {singlet} & 8.90 & 9.04 & 
\multirow{2}*{\textemdash} & 15.05 & 15.19 & 
\multirow{2}*{\textemdash} & 15.28 & \textendash0.09\\ 
(C$_{54}$H$_{18}$) & & & & {\bf triplet} & {\bf 8.81} & {\bf 8.85} & & 
{\bf 14.95} & {\bf 15.00} & & {\bf 15.19} & {\bf \textendash0.19}\\
\noalign{\smallskip}
\hline
\noalign{\smallskip}
Circumovalene &  
\multirow{2}*{5.71} & \multirow{2}*{5.74} & \multirow{2}*{\textemdash} & 
{\bf singlet} & {\bf 8.32} & {\bf 8.34} & \multirow{2}*{\textemdash} & 
{\bf 14.02} & {\bf 14.12} & \multirow{2}*{\textemdash} & 14.07 & +0.05 \\ 
(C$_{66}$H$_{20}$) & & & & triplet & 8.67 & 8.70 & & 14.37 & 14.42 & & 14.44 &
\textendash{} 0.02\\
\end{longtable}}
\end{landscape}

\end{document}